\documentclass[aps,twocolumn,prl,showpacs,superscriptaddress,groupedaddress]{revtex4}

\usepackage{dcolumn}
\usepackage{bm}
\usepackage{color}

\def\ltape{\hbox{\ $<$\hskip -8pt\raise -4pt\hbox{$\sim$}\ }}
\def\gtape{\hbox{\ $>$\hskip -8pt\raise -4pt\hbox{$\sim$}\ }}

   \ifx\pdfoutput\undefined
   \usepackage[dvips]{graphicx}
   \else
   \usepackage[pdftex]{graphicx}
   \fi
   \usepackage{epstopdf}
   \DeclareGraphicsExtensions{.pdf,.gif,.jpg,.eps,.png}
  
   \usepackage{amsmath}
  \usepackage{amssymb}

\pdfoutput=1

\begin{document}


\title{The Critical Role of Collisionless Plasma Energization on the Structure of Relativistic Magnetic Reconnection}

\author{Yi-Hsin~Liu}
\affiliation{Dartmouth College, Hanover, NH 03750}
\author{S. -C.~Lin}
\affiliation{Dartmouth College, Hanover, NH 03750}
\author{M.~Hesse}
\affiliation{University of Bergen, Bergen, Norway}
\affiliation{Southwest Research Institute, San Antonio, TX 78238}
\author{F.~Guo}
\affiliation{Los Alamos National Laboratory, Los Alamos, NM 87545}
\author{X.~Li}
\affiliation{Los Alamos National Laboratory, Los Alamos, NM 87545}
\author{H.~Zhang}
\affiliation{Purdue University, West Lafayette, IN 47907}
\author{S.~Peery}
\affiliation{Dartmouth College, Hanover, NH 03750}

\date{\today}

\begin{abstract}
During magnetically dominated relativistic reconnection, inflowing plasma depletes the initial relativistic pressure at the x-line and collisionless plasma heating inside the diffusion region is insufficient to overcome this loss. The resulting pressure drop causes a collapse at the x-line, essentially a localization mechanism of the diffusion region necessary for fast reconnection. The extension of this low-pressure region further explains the bursty nature of anti-parallel reconnection because a once opened outflow exhaust can also collapse, which repeatedly triggers secondary tearing islands. However, a stable single x-line reconnection can be achieved when an external guide field exists, since the reconnecting magnetic field component rotates out of the reconnection plane at outflows, providing additional magnetic pressure to sustain the opened exhausts. 
\end{abstract}


\maketitle

{\it Introduction--}
The last decade has seen a dramatic surge of interest in the potential role of magnetically dominated reconnection (ratio of the magnetic energy density to the plasma enthalpy density $\sigma\equiv B_0^2/(4\pi w) \gg 1$) in powering strong particle acceleration and high-energy radiation in various astrophysical environments \cite{sironi14a, FGuo14a, FGuo19a, werner16a}, including super-flares in pulsar winds  \cite{Abdo2011,Tavani2011,Uzdensky2011,Arons2012}, accretion disks and jets emanating from rotating compact objects and their merging events \cite{FGuo16a,Zhang2018,Giannios2009,Zhang2011,McKinney2012}. 
Magnetic reconnection breaks and rejoins magnetic field lines inside the {\it diffusion region} that dwells in current sheets. By virtue of the {\it frozen-in} condition between a plasma and magnetic flux outside the diffusion region, a continuous reconnection process inherently involves the transient motion of particles through this diffusion region from the inflow to the outflow areas.
Notably, inside a planar high-$\sigma$ current sheet, the pressure needs to be relativistic to balance the strong upstream magnetic pressure; i.e., $P_{sheet}\approx B_{0}^2/8\pi$.
This balanced pressure during reconnection is often assumed in theoretical models \cite[e.g.,][]{blackman1994,lyubarsky05a,parker57a,uzdensky10a}.
However, it is questionable whether the reconnection diffusion region can provide sufficient thermal heating to sustain this relativistic pressure under a constant inflow of low-pressure (i.e., compared to $B^2_{0}/8\pi$) plasmas. 
Understanding this force balance is critical in determining the structure of the reconnection layer, which ultimately decides particle acceleration during reconnection and its radiation signatures. 

In this letter, we demonstrate that a significant pressure drop occurs at the magnetic x-line, $P_{xline} \ll B_{0}^2/8\pi$, in fully kinetic simulations of high-$\sigma$ magnetic reconnection. We then perform analyses to show that collisionless plasma heating inside the diffusion region is insufficient to sustain a thermal pressure that can balance the strong magnetic pressure far upstream. This plays a key role in determining the geometry of the reconnection layer as it provides a {\it localization mechanism} that limits the length of the diffusion region, and is essential for facilitating fast magnetic reconnection \citep{Biskamp2001} in this regime \footnote{Note that even a similar fast rate $\sim \mathcal{O}(0.1)$ is observed, the localization mechanism can be different in different systems \cite{yhliu18c}.}.
On the other hand, numerical simulations also reveal that relativistic reconnection in the antiparallel geometry is characterized by repetitive bursts of magnetic islands \cite{sironi16a, FGuo15a}, but a more stable single x-line is possible with an external guide field \cite{ball19a, Rowan2019}. We point out that this morphology difference can be explained by the change of the outflow magnetic structure and the pressure balance across the exhaust.

{\it Simulation setup--}
The initial magnetic field ${\bf B}=B_{x0} [\mbox{tanh}(z/\lambda) \hat{\bf x}+ b_g \hat{\bf y}]$. We use electron-positron pairs that have mass $m_i=m_e\equiv m$. Each species has a distribution $f_h \propto \mbox{sech}^2(z/\lambda)\mbox{exp}[-\gamma_d(\gamma_Lmc^2\pm mV_d u_y)/T']$ in the simulation frame, which is a component with a peak density $n'_0$ and temperature $T'$ boosted by a drift velocity $\pm V_d$ in the y-direction for ions and electrons, respectively. In this Letter, the primed quantities are measured in the fluid rest (proper) frame, while the unprimed quantities are measured in the simulation frame unless otherwise specified. Here ${\bf u}=\gamma_L {\bf v}$ is the 4-velocity, $\gamma_L=1/[1-(v/c)^2]^{1/2}$ is the Lorentz factor of a particle, and $\gamma_d \equiv 1/[1-(V_d/c)^2]^{1/2}$. The drift velocity is determined by Amp\'ere's law $cB_{x0}/(4\pi\lambda)=2 e\gamma_d n'_0 V_d $.  
The temperature is determined by the pressure balance $B_{x0}^2/(8\pi)=2 n'_0 T'$.
The resulting density in the simulation frame is $n_0=\gamma_d n'_0$.  In addition, a  non-drifting background component $f_b \propto \mbox{exp}(-\gamma_L m c^2/T_b)$ with a uniform density $n_b$ is included. 
The simulations are performed using VPIC \cite{bowers09a}, which solves the fully relativistic dynamics of particles and electromagnetic fields. 
Densities are normalized by the initial background density $n_b$, time is normalized by the plasma frequency $\omega_{pe}\equiv(4\pi n_b e^2/m_e)^{1/2}$, velocities are normalized by the light speed $c$, and spatial scales are normalized by the inertia length $d_e\equiv c/\omega_{pe}$. Pressures that will be discussed in detail are normalized to $n_b mc^2$. The boundary conditions are periodic in the x-direction, while in the z-direction the field boundary condition is conducting and the particles are reflected at the boundaries. 
The domain size is $L_x\times L_z=768d_e \times 768d_e$ with $6144\times 12288$ cells. There are 100 particles per cell. The half-thickness of the initial sheet is $\lambda=20d_e$, $n_b=n'_0$, $T_b/m_ec^2=0.5$ and $\omega_{pe}/\Omega_{ce}=0.05$ where $\Omega_{ce}\equiv eB_{x0}/(m_e c)$ is a cyclotron frequency.  The magnetization parameter $\sigma\equiv B_0^2/(4\pi w)$ where $w=2n'mc^2+2(\Gamma/(\Gamma-1))n'T'$ with the ratio of specific heats $\Gamma=5/3$. The reconnecting component contributes to $\sigma_{x} \equiv B_{x0}^2/(4\pi w)=(\Omega_{ce}/\omega_{pe})^2/\{2[1+(\Gamma/\Gamma-1)(T_b/m c^2)]\}$, which is $88.9$. In this work, we compare the antiparallel case ($b_g=0$) and a guide field case with $b_g=1$.

\begin{figure}
\includegraphics[width=8cm]{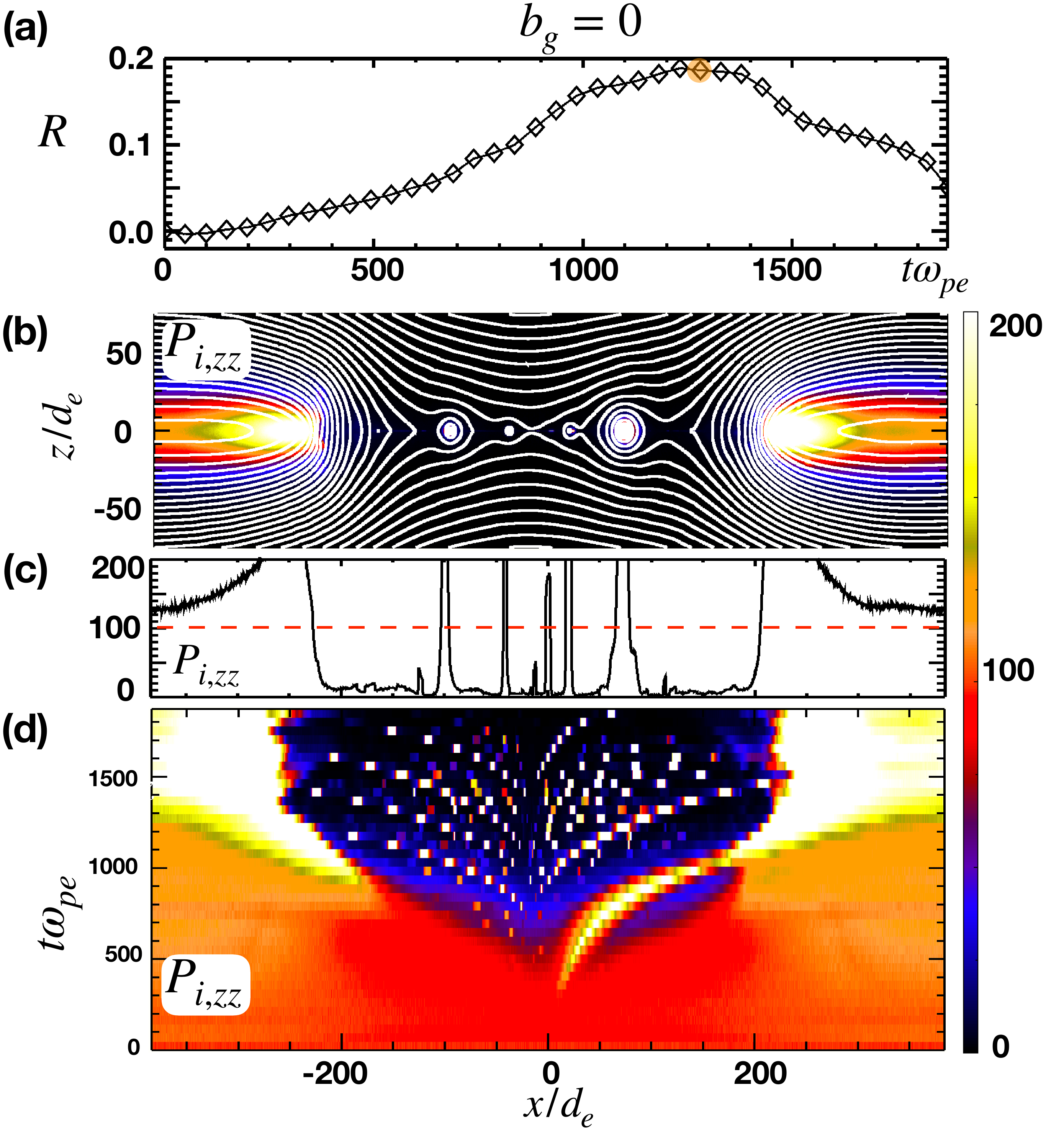} 
\caption {$b_g=0$ case. In (a) the evolution of the reconnection rate $R$. The pressure component $P_{i,zz}$ overlaid with $A_y$ contours at time $1250/\omega_{pe}$ in (b), its cut along $z=0$ in (c) where the red dashed line marks the initial value. The time stack plot of these $z=0$ cuts in (d). Pressures are normalized to $n_bmc^2$ and the color map is caped by value 200.} 
\label{anti}
\end{figure}

\begin{figure}
\includegraphics[width=8cm]{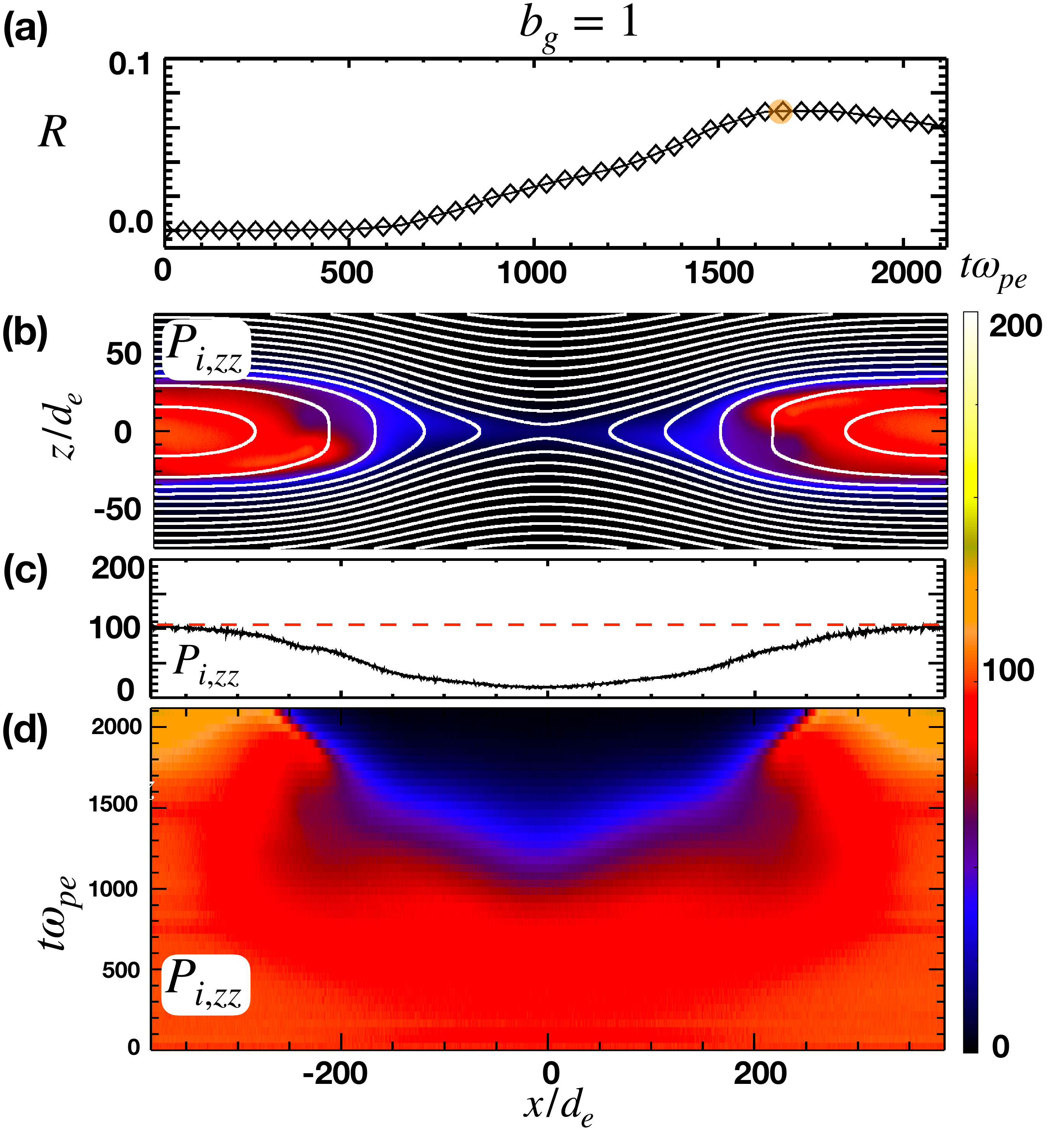} 
\caption {$b_g=1$ case. In (a) the evolution of the reconnection rate $R$. The pressure component $P_{i,zz}$ overlaid with $A_y$ contours at time $1650/\omega_{pe}$ in (b), its cut along $z=0$ in (c) where the red dashed line marks the initial value. The time stack plot of these $z=0$ cuts in (d).} 
\label{bg}
\end{figure}

{\it The pressure depletion at the x-line--}
Fig.~1 shows results in the anti-parallel geometry. Panel (a) shows the evolution of the normalized reconnection rate $R\equiv \partial_t\Psi/B_{x0}V_{Ax0}$, where $\Psi=\mbox{max}(A_y)-\mbox{min}(A_y)$ along $z=0$ and $A_y$ is the y-component of the vector potential. The Alfv\'en speed in the x-direction is $V_{Ax0}=c[\sigma_{x}/(1+\sigma_{x}+\sigma_g)]^{1/2}$ \cite{yhliu15a}. The reconnection rate reaches the typical fast rate of order 0.1 \cite{yhliu17a,cassak17a}. 
The pressure component $P_{i,zz}$, which is responsible for pressure balance across the current sheet, is shown in (b), its cut along $z=0$ in (c), and the time stack plot of $z=0$ cuts in (d). Here we employ Wright and Hadley's \cite{wright75a,hesse07a,zenitani18a} definition of pressure tensor $\tensor P\equiv \int{d^3u{\bf vu}f}-n{\bf VU}$ where ${\bf V}\equiv (1/n)\int{d^3u{\bf v}f}$ and ${\bf U}\equiv (1/n)\int{d^3u{\bf u}f}$. Although the definition of this pressure is not symmetric, it is sufficient to illustrate the pressure change inside the reconnection layer.  
The pronounced feature is a significant drop of pressure (dark area, up to $\times\mathcal{O}(100)$ smaller) at both the x-line and outflow exhausts when the system evolves toward its nonlinear stage, and it is accompanied with the bursty generation of secondary tearing islands. This pressure drop occurs as the inflowing low-pressure plasma from upstream depletes the pressure around the diffusion region. 
For the $b_g=1$ case (Fig. 2), the pressure drop is also evident. An important difference to the antiparallel case is the fact that a stable single x-line and a similar fast rate can be achieved without multiple magnetic islands. 
While not being the focus of this work, interestingly, the thickness of the diffusion region becomes much broader in the $b_g=1$ case likely due to the current starvation effect and incompressibility associated with a guide field \cite{zenitani08a}. 
Note that this pressure depletion and these conclusions hereafter not only apply to Harris-type current sheets, but also to (initially) force-free current sheet \cite{FGuo14a,FGuo15a,yhliu15a}; because the initial magnetic pressure therein will also be expelled out to the downstream in the nonlinear stage.

\begin{figure}
\includegraphics[width=8.5cm]{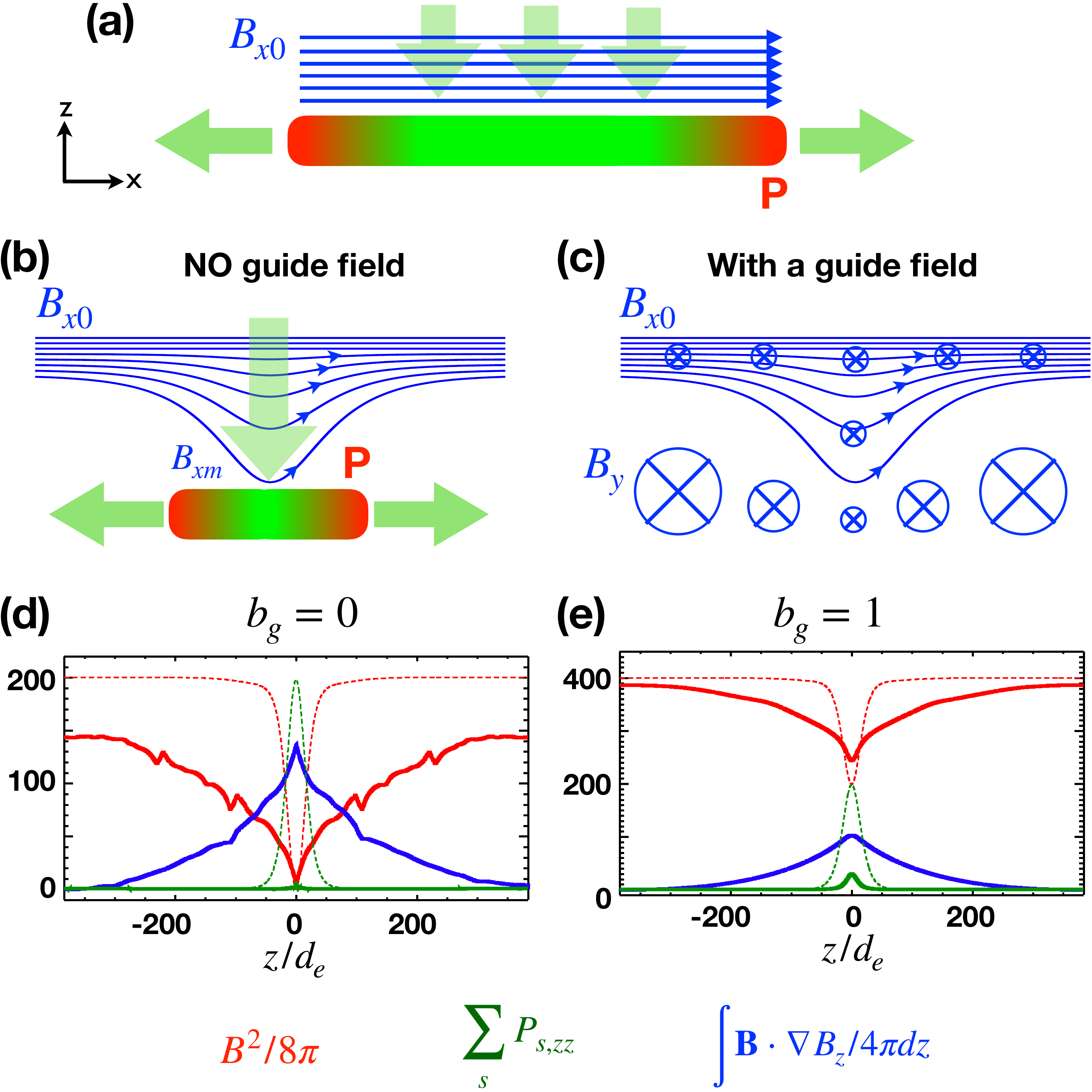} 
\caption {Pressure depletion vs. x-line localization. Green arrows indicate the flow pattern inherent to reconnection. The initial high pressure plasmas in red, the depleted plasma pressure in green. In (a) an elongated diffusion region. In (b) a localized diffusion region. In (c) the effect of the guide field. In (d) the analysis of the force balance (Eq.~(1)) across the primary x-line of the antiparallel case, in (e) for the $b_g=1$ case. Dashed curves show the initial profiles for comparison.} 
\label{localization_mechanism}
\end{figure}

{\it Pressure depletion as a localization mechanism--}
In Fig.~3, we illustrate the need of localization when the thermal pressure drops right at the x-line. During reconnection, the total pressure (i.e., magnetic plus thermal) at the diffusion region is depleted by the inflowing low-pressure plasmas. The red areas of the current sheet in (a) and (b) indicate the original pressure that is high enough to balance the magnetic pressure upstream of the planar current sheet. The green part indicates a low-pressure plasma that flows in from the upstream region. If the pressure depletion cannot be overcome by thermal heating at the diffusion region, then an elongated diffusion region, as shown in (a), is not an option for a steady state solution because the green region will collapse. The only way to restore the force-balance along the inflow is to develop a localized geometry as shown in (b); because the indented upstream magnetic field will invoke a tension force pointing to the upstream, balancing the magnetic pressure gradient force $-\partial_z B_{x0}^2/8\pi\hat{z}$.
Note that the geometry in (b) and (c) with a opened outflow exhaust implies a diffusion region of limited length; i.e., a {\it localized} diffusion region. A more localized diffusion region induces a faster inflow and thus a stronger pressure depletion to localize the diffusion region; i.e., these steps form a dynamical loop of positive feedback.
While one may consider that some other mechanisms, such as secondary tearing modes, localize the diffusion region and deplete the pressure therein accordingly, it is difficult to explain why the pressure inside the entire exhaust is depleted as well, as seen between $x/d_e\in [-200,200]$ in Fig.~1(b)-(d). 

To demonstrate the correlation between the pressure drop and localization, we analyze the force-balance, which can be derived from the momentum equation \cite{hesse07a} $mn_s\partial_t{\bf U}_s+mn_s{\bf V}_s\cdot \nabla {\bf U}_s=-\nabla\cdot{\tensor P}_s+q_sn_s{\bf E}+q_sn_s({\bf V}_s\times{\bf B}/c)$. By summing up the momentum equations of the two species ($s=e,i$), we obtain the force-balance equation
$\nabla B^2/8\pi+\sum_s \nabla\cdot {\tensor P}_s-{\bf B\cdot\nabla{\bf B}}/4\pi+\sum_s mn_s{\bf V}_s\cdot\nabla{\bf U}_s+\sum_s mn_s\partial_t{\bf U}_s-\sum_s q_s n_s {\bf E}+\partial_t{\bf E}/(4\pi c)=0$. Across the x-line along the inflow (z-) direction, the dominant terms are integrated to give
\begin{equation}
\frac{B^2}{8\pi}+\sum_s P_{s,zz}-\int^z_{-L_z/2}{\frac{{\bf B}\cdot\nabla B_z}{4\pi}}dz'\simeq const.
\end{equation}
As shown in Fig.~3(d), the initial (dashed curves) relativistically hot $\sum_s P_{s,zz}$ (green) can balance the upstream $B^2/8\pi$ (red). Later (solid curves), $\sum_s P_{s,zz}$ drops significantly and the only term that can balance $B^2/8\pi$ is the tension force (blue) pointing to the upstream. This captures the effect of the indenting upstream magnetic field illustrated in (b) and (c), essentially the localization of the diffusion region \cite{yhliu17a}. A similar balance is observed with a guide field in (e), but an important difference at the outflow exhaust will be discussed later.



\begin{figure}
\includegraphics[width=8.5cm]{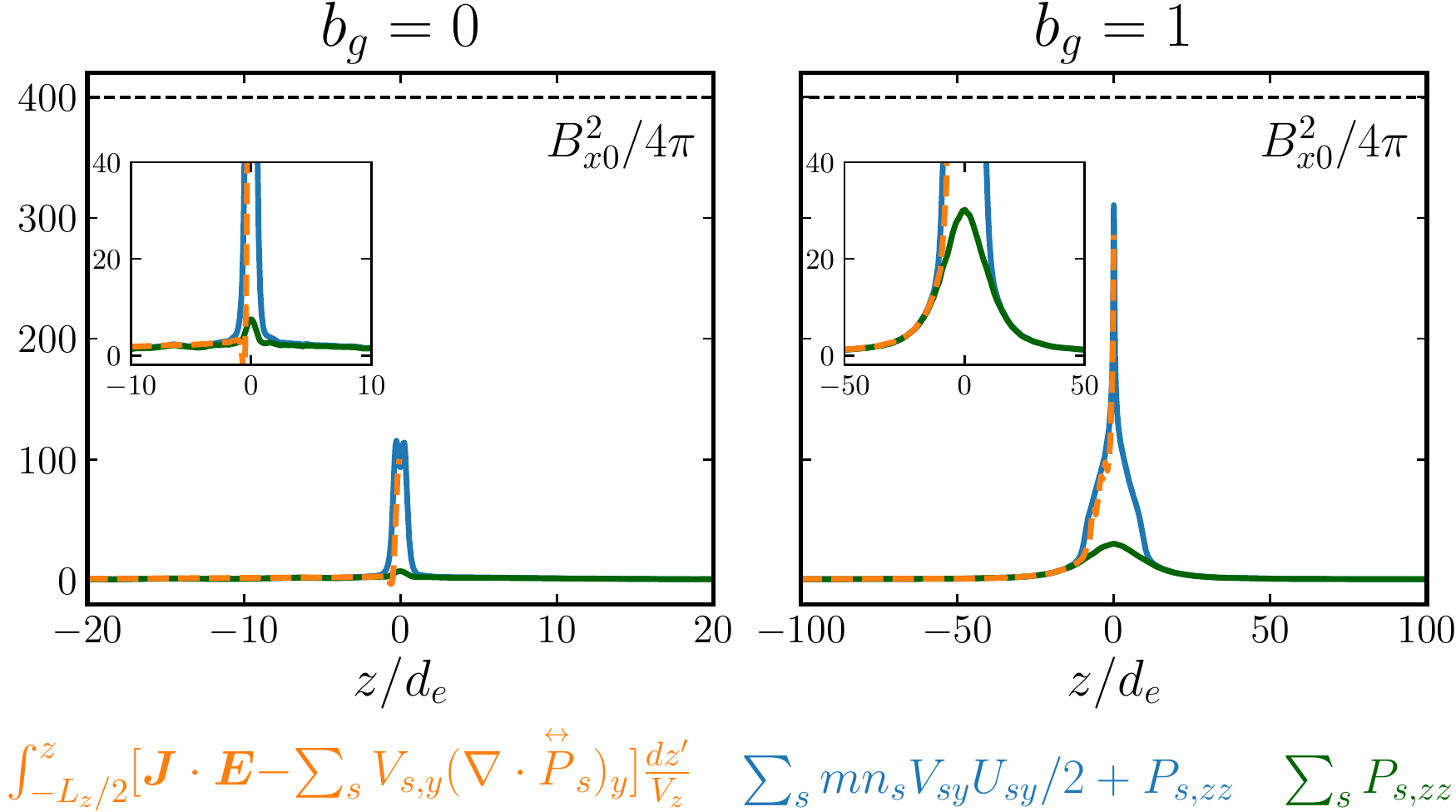} 
\caption {Heating efficiency (Eq.~(2)) analysis across the primary x-line. The $b_g=0$ case ($b_g=1$ case) in the left (right). The insets zoom in near the x-line to better show $\sum_s P_{s,zz}$ profiles.} 
\label{heating_rate}
\end{figure}

{\it The thermal heating efficiency toward the x-line--}
If $P_{zz}$ at the x-line is smaller than $B_{x0}^2/8\pi$, it leads to a localization needed for fast reconnection. This condition holds regardless of the initial thickness ($2\delta$) or the profile of the current sheet. The question is then why the pressure at the x-line drops significantly. We address this question by examining the heating process along with the inflowing plasma.
Per Poynting's theorem, ${\bf J}\cdot{\bf E}$ measures the energy conversion rate from electromagnetic energy to plasma energies.
Dotting the momentum equations with ${\bf V}_s$ then summing up species, we get 
${\bf J}\cdot {\bf E}=\sum_s mn_s{\bf V}_s\cdot ({\bf V}_s\cdot\nabla){\bf U}_s+\sum_s {\bf V}_s\cdot(\nabla\cdot {\tensor P}_s)+\sum_s mn_s {\bf V}_s\cdot\partial_t{\bf U}_s$. Integrating the energy gain of plasmas along its path (at $x=0$) toward the x-line, we find the dominant terms in the nonlinear state,
\begin{equation}
\begin{split}
\int{({\bf J}\cdot {\bf E}) dt}=\int_{-L_z/2}^z{({\bf J}\cdot {\bf E}) \frac{dz'}{V_z}}\\
\simeq \sum_s mn_s\frac{V_{s,y}U_{s,y}}{2}+\sum_s P_{s,zz}+\int \sum_s V_{s,y}(\nabla\cdot {\tensor P}_s)_y\frac{dz'}{V_z}.
\end{split}
\label{heating}
\end{equation}
This integral has an apparent singularity near the vicinity of the x-line where $V_z\rightarrow 0$, which exactly arises from the last term of RHS; i.e., because $E_y=(1/q_sn_s)(\nabla\cdot{\tensor P}_s)_y$ \cite{hesse11a} right at the x-line, so that ${\bf J}\cdot {\bf E}\simeq \sum_sV_{s,y}(\nabla\cdot{\tensor P}_s)_y$. 
We remove the contribution from this term in the integral \footnote{ Note that a larger $(\nabla\cdot{\tensor P}_s)_y$ does not help with the pressure balance in the z-direction even in the non-relativistic limit where $\tensor{P}_s$ is symmetric, because $\partial_y P_{s,yz}=0$ in 2D.} 
and plot it as orange curves in Fig.~4 for both the $b_g=0$ and $b_g=1$ cases. We see that these orange curves follow well the profiles of $\sum_s mn_sV_{s,y}U_{s,y}/2+\sum_sP_{s,zz}$ in light blue. In the insets, $\sum_s P_{s,zz}$ profiles (in green) are blown up to better show the variation. In both cases, the magnetic energy is mostly converted to the bulk flow kinetic energy in the current (y-) direction, while only a relatively small portion to the thermal pressure in the z-direction, $\sum_s P_{s,zz}$.

On the other hand, a rough estimation suggests that the energy conversion $\int{({\bf J}\cdot {\bf E})dt}\sim J_y E_y \Delta t \sim (c/4\pi)(B_{x0}/\delta) (V_{in} B_{x0}/c) (\delta/V_{in})\sim \mathcal{O}(B_{x0}^2/4\pi)$. 
Here $\Delta t\sim \delta/V_{in}$ is used to estimate the transient time-scale \cite{hesse11a}; i.e., the average time spent by a particle in the diffusion region. The peak value of plasma energy gain (in light blue) is limited by $\mathcal{O}(B_{x0}^2/4\pi)=400$ (dashed lines) as shown in Fig.~4. From these observations, we conclude that if most energy is converted to the bulk kinetic energy of the current carrier, then right at the x-line $\sum_sP_{s,zz} < B^2_{x0}/8\pi$; the diffusion region needs to be localized. 
It is interesting to remark that this kinetic description is different from that of resistive-MHD models. In MHD, inflowing plasma does not need to be turned into current carriers and no energy is required to sustain the current. Thus, such localization mechanisms may be absent in resistive-MHD.

\begin{figure}
\includegraphics[width=8 cm]{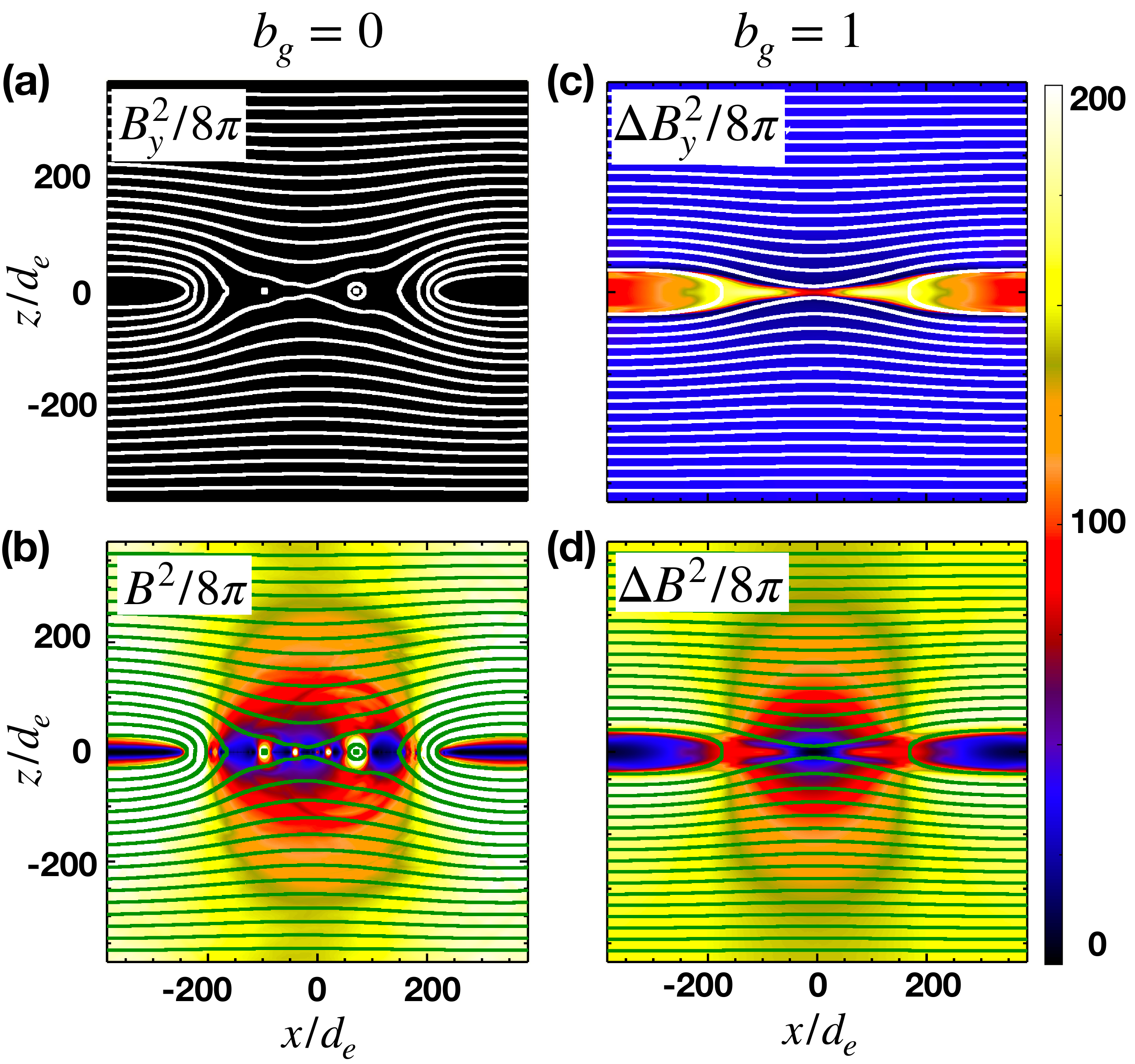} 
\caption {The $b_g=0$ case ($b_g=1$ cases) in the left (right) column. Top and bottom rows show the out-of-plane magnetic pressure $B^2_y/8\pi$ and the total magnetic pressure $B^2/8\pi$ overlaid with $A_y$ contours, respectively.  In (c) and (d), quantity $\Delta Q \equiv Q-\mbox{min}(Q)$. The color map In (b) is caped by value 200. The concentric $B^2/8\pi$ dip around the x-line  best illustrates the localization.} 
\label{comparison}
\end{figure}

{\it Bursty multiple x-lines vs. stable single x-line--}
The preferential pressure depletion right at the x-line tends to localize the diffusion region. For an opened outflow exhaust to be {\it stable}, it also requires a balanced pressure across the exhaust; i.e., it needs the high-pressure (red) parts of the current sheet in Fig.~3(b).
Although the plasma will be further heated while being accelerated into outflow exhausts, the exhaust heating in the anti-parallel case (Fig.~1), however, also appears unable to bring the plasma pressure back to the original value. A once opened exhaust will thus collapse into a thin current sheet, until it triggers copious fast growing secondary tearing modes, forming competing multiple x-lines. The growth of magnetic islands helps establish the localization (Fig.~3(b)) locally for each individual x-line, but those islands can be expelled out by the primary outflows from a primary x-line (near the center of simulation domain). Thus, the generation of magnetic islands inside the reconnection layer is bursty and repetitive, as clearly seen in Fig.~1(d). 
One may argue that these tearing modes are essential for the localization, but they are in fact secondary effects immersed inside the large-scale localization (i.e., concentric darker area in Fig.~5(b)) caused by the pressure depletion. 
This aspect becomes clearer in guide field reconnection where secondary tearing islands can be avoided but the system still achieves a localized geometry. Due to the symmetry, an out-of-plane magnetic field is not generated in the anti-parallel case as shown in Fig.~5(a). However, with a guide field the reconnecting field, once reconnects, can simply rotate to the out-of-plane direction and provide the (magnetic) pressure needed for supporting opened outflow exhausts. This is seen in Fig.~5(c) and (d) along the outflow, and the idea is illustrated in Fig.~3(c) where a larger $\otimes$ symbol indicates a stronger out-of-plane field. This structure is consistent with Petschek's solution of outflow structures; in the anti-parallel case, the outflow exhaust is bounded by a pair of co-planar (i.e., no $B_y$ at downstream) slow shocks \cite{petschek64a,lin93a,lyubarsky05a}, which turns into a pair of rotational discontinuities (that preserves the magnetic pressure) in the guide field case [e.g., \cite{levy64a,lin93a,lyubarsky05a,yhliu11b}]. This additional source of magnetic pressure (Fig.~5(c)) inside the exhaust makes a stable single x-line reconnection possible, in sharp contrast to the bursty antiparallel case (Fig.~5(b)).


{\it Summary and Discussion--}
In strongly magnetized plasmas, there is an intriguing linkage between the heating efficiency inside the reconnection diffusion region and its localization mechanism, that is needed for fast reconnection. 
We analyze the force balance across the diffusion region of relativistic reconnection with and without a guide field. For both cases, significant pressure drops from the original equilibrium value are observed at the x-line. The inflowing plasma gains mostly the bulk kinetic energy in the out-of-plane direction, while only a small fraction of magnetic energy is converted to build up the pressure in the inflow direction; this may reflect the difficulty of thermal heating (compared to the bulk acceleration) in collisionless plasmas. Meanwhile, the overall energy conversion seen by the plasma flowing into the x-line is limited due to its transient time-scale within the diffusion region. Thus we conclude that the thermal heating inside the diffusion region is insufficient to overcome the pressure depletion by the continuous inflowing low-pressure plasma, making $P_{zz} < B^2_{x0}/8\pi$ right at the x-line. This pressure drop localizes the diffusion region.
Contrary to the common perception on the role of x-line heating, we argue that an insufficient thermal heating at reconnection x-line, in fact, provides a key localization mechanism necessary for fast reconnection. Given some degree of localization, the system can easily reach a state with a reconnection rate close to value $\sim \mathcal{O}(0.1)$ \cite{yhliu17a}.

Radiative cooling could further reduce the thermal pressure at the x-line, enhancing the localization. It may also trigger more secondary tearing islands during antiparallel reconnection because the low-pressure region can extend to farther downstream.
In nature it is more common to have a finite guide field, and the extra magnetic pressure provided by the field rotation helps support the opening of reconnection exhausts, enabling a stable single x-line.
Interestingly, in electron-proton plasmas the Hall quadrupole field also arises from the rotation of the reconnecting magnetic field \cite{mandt94a,drake08a}, and it may play a similar role in providing an additional pressure to support the opened outflow exhaust \footnote{Note that the Hall quadrupole field does not ease the pressure depletion at the x-line, since it vanishes right at the x-line.}. This aspect deserves a thorough study both in the high-$\sigma$ and in the non-relativistic low-$\beta$ regimes. \\




\acknowledgments
Y.-H. L is grateful for supports from grants NSF-DoE 1902867 and NASA MMS 80NSSC18K0289. F. G. acknowledges support from DOE through the LDRD program at LANL and DoE/OFES support to LANL, and NASA ATP program through grant NNH17AE68I. X. L.'s contribution is in part supported by NASA under grant NNH16AC60I. Simulations were performed at the National Center for Computational Sciences at ORNL and with LANL institutional computing. 


\begin{thebibliography}{38}
\expandafter\ifx\csname natexlab\endcsname\relax\def\natexlab#1{#1}\fi
\expandafter\ifx\csname bibnamefont\endcsname\relax
  \def\bibnamefont#1{#1}\fi
\expandafter\ifx\csname bibfnamefont\endcsname\relax
  \def\bibfnamefont#1{#1}\fi
\expandafter\ifx\csname citenamefont\endcsname\relax
  \def\citenamefont#1{#1}\fi
\expandafter\ifx\csname url\endcsname\relax
  \def\url#1{\texttt{#1}}\fi
\expandafter\ifx\csname urlprefix\endcsname\relax\def\urlprefix{URL }\fi
\providecommand{\bibinfo}[2]{#2}
\providecommand{\eprint}[2][]{\url{#2}}

\bibitem[{\citenamefont{Sironi and Spitkovsky}(2014)}]{sironi14a}
\bibinfo{author}{\bibfnamefont{L.}~\bibnamefont{Sironi}} \bibnamefont{and}
  \bibinfo{author}{\bibfnamefont{A.}~\bibnamefont{Spitkovsky}},
  \bibinfo{journal}{Astrophys. J.} \textbf{\bibinfo{volume}{783}},
  \bibinfo{pages}{L21} (\bibinfo{year}{2014}).

\bibitem[{\citenamefont{Guo et~al.}(2014)\citenamefont{Guo, Li, Daughton, and
  {Yi-Hsin Liu}}}]{FGuo14a}
\bibinfo{author}{\bibfnamefont{F.}~\bibnamefont{Guo}},
  \bibinfo{author}{\bibfnamefont{H.}~\bibnamefont{Li}},
  \bibinfo{author}{\bibfnamefont{W.}~\bibnamefont{Daughton}}, \bibnamefont{and}
  \bibinfo{author}{\bibnamefont{{Yi-Hsin Liu}}}, \bibinfo{journal}{Phys. Rev.
  Lett.} \textbf{\bibinfo{volume}{113}}, \bibinfo{pages}{155005}
  (\bibinfo{year}{2014}).

\bibitem[{\citenamefont{Guo et~al.}(2019)\citenamefont{Guo, Li, Daughton,
  Kilian, Li, Liu, Yan, and Ma}}]{FGuo19a}
\bibinfo{author}{\bibfnamefont{F.}~\bibnamefont{Guo}},
  \bibinfo{author}{\bibfnamefont{X.}~\bibnamefont{Li}},
  \bibinfo{author}{\bibfnamefont{W.}~\bibnamefont{Daughton}},
  \bibinfo{author}{\bibfnamefont{P.}~\bibnamefont{Kilian}},
  \bibinfo{author}{\bibfnamefont{H.}~\bibnamefont{Li}},
  \bibinfo{author}{\bibfnamefont{Y.-H.} \bibnamefont{Liu}},
  \bibinfo{author}{\bibfnamefont{W.}~\bibnamefont{Yan}}, \bibnamefont{and}
  \bibinfo{author}{\bibfnamefont{D.}~\bibnamefont{Ma}},
  \bibinfo{journal}{Astrophys. J. Lett.} \textbf{\bibinfo{volume}{879}},
  \bibinfo{pages}{23} (\bibinfo{year}{2019}).

\bibitem[{\citenamefont{Werner et~al.}(2016)\citenamefont{Werner, Uzdensky,
  Cerutti, Nalewajko, and Begelman}}]{werner16a}
\bibinfo{author}{\bibfnamefont{G.~R.} \bibnamefont{Werner}},
  \bibinfo{author}{\bibfnamefont{D.~A.} \bibnamefont{Uzdensky}},
  \bibinfo{author}{\bibfnamefont{B.}~\bibnamefont{Cerutti}},
  \bibinfo{author}{\bibfnamefont{K.}~\bibnamefont{Nalewajko}},
  \bibnamefont{and} \bibinfo{author}{\bibfnamefont{M.~C.}
  \bibnamefont{Begelman}}, \bibinfo{journal}{Astrophys. J. Lett.}
  \textbf{\bibinfo{volume}{816}}, \bibinfo{pages}{L8} (\bibinfo{year}{2016}).

\bibitem[{\citenamefont{{Abdo} et~al.}(2011)\citenamefont{{Abdo}, {Ackermann},
  {Ajello}, {Allafort}, {Baldini}, {Ballet}, {Barbiellini}, {Bastieri},
  {Bechtol}, {Bellazzini} et~al.}}]{Abdo2011}
\bibinfo{author}{\bibfnamefont{A.~A.} \bibnamefont{{Abdo}}},
  \bibinfo{author}{\bibfnamefont{M.}~\bibnamefont{{Ackermann}}},
  \bibinfo{author}{\bibfnamefont{M.}~\bibnamefont{{Ajello}}},
  \bibinfo{author}{\bibfnamefont{A.}~\bibnamefont{{Allafort}}},
  \bibinfo{author}{\bibfnamefont{L.}~\bibnamefont{{Baldini}}},
  \bibinfo{author}{\bibfnamefont{J.}~\bibnamefont{{Ballet}}},
  \bibinfo{author}{\bibfnamefont{G.}~\bibnamefont{{Barbiellini}}},
  \bibinfo{author}{\bibfnamefont{D.}~\bibnamefont{{Bastieri}}},
  \bibinfo{author}{\bibfnamefont{K.}~\bibnamefont{{Bechtol}}},
  \bibinfo{author}{\bibfnamefont{R.}~\bibnamefont{{Bellazzini}}},
  \bibnamefont{et~al.}, \bibinfo{journal}{Science}
  \textbf{\bibinfo{volume}{331}}, \bibinfo{pages}{739} (\bibinfo{year}{2011}),
  \eprint{1011.3855}.

\bibitem[{\citenamefont{{Tavani} et~al.}(2011)\citenamefont{{Tavani},
  {Bulgarelli}, {Vittorini}, {Pellizzoni}, {Striani}, {Caraveo}, {Weisskopf},
  {Tennant}, {Pucella}, {Trois} et~al.}}]{Tavani2011}
\bibinfo{author}{\bibfnamefont{M.}~\bibnamefont{{Tavani}}},
  \bibinfo{author}{\bibfnamefont{A.}~\bibnamefont{{Bulgarelli}}},
  \bibinfo{author}{\bibfnamefont{V.}~\bibnamefont{{Vittorini}}},
  \bibinfo{author}{\bibfnamefont{A.}~\bibnamefont{{Pellizzoni}}},
  \bibinfo{author}{\bibfnamefont{E.}~\bibnamefont{{Striani}}},
  \bibinfo{author}{\bibfnamefont{P.}~\bibnamefont{{Caraveo}}},
  \bibinfo{author}{\bibfnamefont{M.~C.} \bibnamefont{{Weisskopf}}},
  \bibinfo{author}{\bibfnamefont{A.}~\bibnamefont{{Tennant}}},
  \bibinfo{author}{\bibfnamefont{G.}~\bibnamefont{{Pucella}}},
  \bibinfo{author}{\bibfnamefont{A.}~\bibnamefont{{Trois}}},
  \bibnamefont{et~al.}, \bibinfo{journal}{Science}
  \textbf{\bibinfo{volume}{331}}, \bibinfo{pages}{736} (\bibinfo{year}{2011}),
  \eprint{1101.2311}.

\bibitem[{\citenamefont{{Uzdensky} et~al.}(2011)\citenamefont{{Uzdensky},
  {Cerutti}, and Begelman}}]{Uzdensky2011}
\bibinfo{author}{\bibfnamefont{D.~A.} \bibnamefont{{Uzdensky}}},
  \bibinfo{author}{\bibfnamefont{B.}~\bibnamefont{{Cerutti}}},
  \bibnamefont{and} \bibinfo{author}{\bibfnamefont{M.~C.}
  \bibnamefont{Begelman}}, \bibinfo{journal}{ApjL}
  \textbf{\bibinfo{volume}{737}}, \bibinfo{eid}{L40} (\bibinfo{year}{2011}),
  \eprint{1105.0942}.

\bibitem[{\citenamefont{Arons}(2012)}]{Arons2012}
\bibinfo{author}{\bibfnamefont{J.}~\bibnamefont{Arons}}, \bibinfo{journal}{SSR}
  \textbf{\bibinfo{volume}{173}}, \bibinfo{pages}{341} (\bibinfo{year}{2012}),
  \eprint{1208.5787}.

\bibitem[{\citenamefont{Guo et~al.}(2016)\citenamefont{Guo, Li, Li, Daughton,
  Zhang, {Lloyd-Ronning}, {Yi-Hsin Liu}, Zhang, and Deng}}]{FGuo16a}
\bibinfo{author}{\bibfnamefont{F.}~\bibnamefont{Guo}},
  \bibinfo{author}{\bibfnamefont{X.}~\bibnamefont{Li}},
  \bibinfo{author}{\bibfnamefont{H.}~\bibnamefont{Li}},
  \bibinfo{author}{\bibfnamefont{W.}~\bibnamefont{Daughton}},
  \bibinfo{author}{\bibfnamefont{B.}~\bibnamefont{Zhang}},
  \bibinfo{author}{\bibfnamefont{N.}~\bibnamefont{{Lloyd-Ronning}}},
  \bibinfo{author}{\bibnamefont{{Yi-Hsin Liu}}},
  \bibinfo{author}{\bibfnamefont{H.}~\bibnamefont{Zhang}}, \bibnamefont{and}
  \bibinfo{author}{\bibfnamefont{W.}~\bibnamefont{Deng}},
  \bibinfo{journal}{Astrophys. J. Lett.} \textbf{\bibinfo{volume}{818}},
  \bibinfo{pages}{L9} (\bibinfo{year}{2016}).

\bibitem[{\citenamefont{{Zhang} et~al.}(2018)\citenamefont{{Zhang}, {Li},
  {Guo}, and {Giannios}}}]{Zhang2018}
\bibinfo{author}{\bibfnamefont{H.}~\bibnamefont{{Zhang}}},
  \bibinfo{author}{\bibfnamefont{X.}~\bibnamefont{{Li}}},
  \bibinfo{author}{\bibfnamefont{F.}~\bibnamefont{{Guo}}}, \bibnamefont{and}
  \bibinfo{author}{\bibfnamefont{D.}~\bibnamefont{{Giannios}}},
  \bibinfo{journal}{ApjL} \textbf{\bibinfo{volume}{862}}, \bibinfo{eid}{L25}
  (\bibinfo{year}{2018}), \eprint{1807.08420}.

\bibitem[{\citenamefont{{Giannios} et~al.}(2009)\citenamefont{{Giannios},
  {Uzdensky}, and Begelman}}]{Giannios2009}
\bibinfo{author}{\bibfnamefont{D.}~\bibnamefont{{Giannios}}},
  \bibinfo{author}{\bibfnamefont{D.~A.} \bibnamefont{{Uzdensky}}},
  \bibnamefont{and} \bibinfo{author}{\bibfnamefont{M.~C.}
  \bibnamefont{Begelman}}, \bibinfo{journal}{MNRAS}
  \textbf{\bibinfo{volume}{395}}, \bibinfo{pages}{L29} (\bibinfo{year}{2009}),
  \eprint{0901.1877}.

\bibitem[{\citenamefont{{Zhang} and {Yan}}(2011)}]{Zhang2011}
\bibinfo{author}{\bibfnamefont{B.}~\bibnamefont{{Zhang}}} \bibnamefont{and}
  \bibinfo{author}{\bibfnamefont{H.}~\bibnamefont{{Yan}}},
  \bibinfo{journal}{\apj} \textbf{\bibinfo{volume}{726}}, \bibinfo{eid}{90}
  (\bibinfo{year}{2011}), \eprint{1011.1197}.

\bibitem[{\citenamefont{{McKinney} and {Uzdensky}}(2012)}]{McKinney2012}
\bibinfo{author}{\bibfnamefont{J.~C.} \bibnamefont{{McKinney}}}
  \bibnamefont{and} \bibinfo{author}{\bibfnamefont{D.~A.}
  \bibnamefont{{Uzdensky}}}, \bibinfo{journal}{MNRAS}
  \textbf{\bibinfo{volume}{419}}, \bibinfo{pages}{573} (\bibinfo{year}{2012}),
  \eprint{1011.1904}.

\bibitem[{\citenamefont{{Blackman} and {Field}}(1994)}]{blackman1994}
\bibinfo{author}{\bibfnamefont{E.~G.} \bibnamefont{{Blackman}}}
  \bibnamefont{and} \bibinfo{author}{\bibfnamefont{G.~B.}
  \bibnamefont{{Field}}}, \bibinfo{journal}{Physical Review Letters}
  \textbf{\bibinfo{volume}{72}}, \bibinfo{pages}{494} (\bibinfo{year}{1994}).

\bibitem[{\citenamefont{Lyubarsky}(2005)}]{lyubarsky05a}
\bibinfo{author}{\bibfnamefont{Y.~E.} \bibnamefont{Lyubarsky}},
  \bibinfo{journal}{MNRAS} \textbf{\bibinfo{volume}{358}}, \bibinfo{pages}{113}
  (\bibinfo{year}{2005}).

\bibitem[{\citenamefont{Parker}(1957)}]{parker57a}
\bibinfo{author}{\bibfnamefont{E.~N.} \bibnamefont{Parker}},
  \bibinfo{journal}{J. Geophys. Res.} \textbf{\bibinfo{volume}{62}},
  \bibinfo{pages}{509} (\bibinfo{year}{1957}).

\bibitem[{\citenamefont{Uzdensky et~al.}(2010)\citenamefont{Uzdensky, Loureiro,
  and Schekochinhin}}]{uzdensky10a}
\bibinfo{author}{\bibfnamefont{D.~A.} \bibnamefont{Uzdensky}},
  \bibinfo{author}{\bibfnamefont{N.~F.} \bibnamefont{Loureiro}},
  \bibnamefont{and} \bibinfo{author}{\bibfnamefont{A.~A.}
  \bibnamefont{Schekochinhin}}, \bibinfo{journal}{Phys. Rev. Lett.}
  \textbf{\bibinfo{volume}{105}}, \bibinfo{pages}{235002}
  (\bibinfo{year}{2010}).

\bibitem[{\citenamefont{{Biskamp} and {Schwarz}}(2001)}]{Biskamp2001}
\bibinfo{author}{\bibfnamefont{D.}~\bibnamefont{{Biskamp}}} \bibnamefont{and}
  \bibinfo{author}{\bibfnamefont{E.}~\bibnamefont{{Schwarz}}},
  \bibinfo{journal}{Physics of Plasmas} \textbf{\bibinfo{volume}{8}},
  \bibinfo{pages}{4729} (\bibinfo{year}{2001}).

\bibitem[{\citenamefont{Sironi et~al.}(2016)\citenamefont{Sironi, Giannios, and
  Petropoulou}}]{sironi16a}
\bibinfo{author}{\bibfnamefont{L.}~\bibnamefont{Sironi}},
  \bibinfo{author}{\bibfnamefont{D.}~\bibnamefont{Giannios}}, \bibnamefont{and}
  \bibinfo{author}{\bibfnamefont{M.}~\bibnamefont{Petropoulou}},
  \bibinfo{journal}{MNRAS} \textbf{\bibinfo{volume}{462}}, \bibinfo{pages}{48}
  (\bibinfo{year}{2016}).

\bibitem[{\citenamefont{{Guo} et~al.}(2015)\citenamefont{{Guo}, {Liu},
  {Daughton}, and {Li}}}]{FGuo15a}
\bibinfo{author}{\bibfnamefont{F.}~\bibnamefont{{Guo}}},
  \bibinfo{author}{\bibfnamefont{Y.-H.} \bibnamefont{{Liu}}},
  \bibinfo{author}{\bibfnamefont{W.}~\bibnamefont{{Daughton}}},
  \bibnamefont{and} \bibinfo{author}{\bibfnamefont{H.}~\bibnamefont{{Li}}},
  \bibinfo{journal}{\apj} \textbf{\bibinfo{volume}{806}}, \bibinfo{eid}{167}
  (\bibinfo{year}{2015}), \eprint{1504.02193}.

\bibitem[{\citenamefont{Ball et~al.}(2019)\citenamefont{Ball, Sironi, and
  Ozel}}]{ball19a}
\bibinfo{author}{\bibfnamefont{D.}~\bibnamefont{Ball}},
  \bibinfo{author}{\bibfnamefont{L.}~\bibnamefont{Sironi}}, \bibnamefont{and}
  \bibinfo{author}{\bibfnamefont{F.}~\bibnamefont{Ozel}},
  \bibinfo{journal}{Astrophys. J.} \textbf{\bibinfo{volume}{884}},
  \bibinfo{pages}{57} (\bibinfo{year}{2019}).

\bibitem[{\citenamefont{{Rowan} et~al.}(2019)\citenamefont{{Rowan}, {Sironi},
  and {Narayan}}}]{Rowan2019}
\bibinfo{author}{\bibfnamefont{M.~E.} \bibnamefont{{Rowan}}},
  \bibinfo{author}{\bibfnamefont{L.}~\bibnamefont{{Sironi}}}, \bibnamefont{and}
  \bibinfo{author}{\bibfnamefont{R.}~\bibnamefont{{Narayan}}},
  \bibinfo{journal}{\apj} \textbf{\bibinfo{volume}{873}}, \bibinfo{eid}{2}
  (\bibinfo{year}{2019}), \eprint{1901.05438}.

\bibitem[{\citenamefont{Bowers et~al.}(2009)\citenamefont{Bowers, Albright,
  Yin, Daughton, Roytershteyn, Bergen, and Kwan}}]{bowers09a}
\bibinfo{author}{\bibfnamefont{K.}~\bibnamefont{Bowers}},
  \bibinfo{author}{\bibfnamefont{B.}~\bibnamefont{Albright}},
  \bibinfo{author}{\bibfnamefont{L.}~\bibnamefont{Yin}},
  \bibinfo{author}{\bibfnamefont{W.}~\bibnamefont{Daughton}},
  \bibinfo{author}{\bibfnamefont{V.}~\bibnamefont{Roytershteyn}},
  \bibinfo{author}{\bibfnamefont{B.}~\bibnamefont{Bergen}}, \bibnamefont{and}
  \bibinfo{author}{\bibfnamefont{T.}~\bibnamefont{Kwan}},
  \bibinfo{journal}{Journal of Physics: Conference Series}
  \textbf{\bibinfo{volume}{180}}, \bibinfo{pages}{012055}
  (\bibinfo{year}{2009}).

\bibitem[{\citenamefont{{Yi-Hsin Liu} et~al.}(2015)\citenamefont{{Yi-Hsin Liu},
  Guo, Daughton, Li, and Hesse}}]{yhliu15a}
\bibinfo{author}{\bibnamefont{{Yi-Hsin Liu}}},
  \bibinfo{author}{\bibfnamefont{F.}~\bibnamefont{Guo}},
  \bibinfo{author}{\bibfnamefont{W.}~\bibnamefont{Daughton}},
  \bibinfo{author}{\bibfnamefont{H.}~\bibnamefont{Li}}, \bibnamefont{and}
  \bibinfo{author}{\bibfnamefont{M.}~\bibnamefont{Hesse}},
  \bibinfo{journal}{Phys. Rev. Lett.} \textbf{\bibinfo{volume}{114}},
  \bibinfo{pages}{095002} (\bibinfo{year}{2015}).

\bibitem[{\citenamefont{Liu et~al.}(2017)\citenamefont{Liu, Hesse, Guo,
  Daughton, Li, Cassak, and Shay}}]{yhliu17a}
\bibinfo{author}{\bibfnamefont{Y.-H.} \bibnamefont{Liu}},
  \bibinfo{author}{\bibfnamefont{M.}~\bibnamefont{Hesse}},
  \bibinfo{author}{\bibfnamefont{F.}~\bibnamefont{Guo}},
  \bibinfo{author}{\bibfnamefont{W.}~\bibnamefont{Daughton}},
  \bibinfo{author}{\bibfnamefont{H.}~\bibnamefont{Li}},
  \bibinfo{author}{\bibfnamefont{P.~A.} \bibnamefont{Cassak}},
  \bibnamefont{and} \bibinfo{author}{\bibfnamefont{M.~A.} \bibnamefont{Shay}},
  \bibinfo{journal}{Phys. Rev. Lett.} \textbf{\bibinfo{volume}{118}},
  \bibinfo{pages}{085101} (\bibinfo{year}{2017}).

\bibitem[{\citenamefont{Cassak et~al.}(2017)\citenamefont{Cassak, {Yi-Hsin
  Liu}, and Shay}}]{cassak17a}
\bibinfo{author}{\bibfnamefont{P.~A.} \bibnamefont{Cassak}},
  \bibinfo{author}{\bibnamefont{{Yi-Hsin Liu}}}, \bibnamefont{and}
  \bibinfo{author}{\bibfnamefont{M.~A.} \bibnamefont{Shay}},
  \bibinfo{journal}{J. Plasma Phys.} \textbf{\bibinfo{volume}{83}},
  \bibinfo{pages}{715830501} (\bibinfo{year}{2017}).

\bibitem[{\citenamefont{Wright and Hadley}(1975)}]{wright75a}
\bibinfo{author}{\bibfnamefont{T.~P.} \bibnamefont{Wright}} \bibnamefont{and}
  \bibinfo{author}{\bibfnamefont{G.~R.} \bibnamefont{Hadley}},
  \bibinfo{journal}{Phys. Rev. A} \textbf{\bibinfo{volume}{12}},
  \bibinfo{pages}{686} (\bibinfo{year}{1975}).

\bibitem[{\citenamefont{Hesse and Zenitani}(2007)}]{hesse07a}
\bibinfo{author}{\bibfnamefont{M.}~\bibnamefont{Hesse}} \bibnamefont{and}
  \bibinfo{author}{\bibfnamefont{S.}~\bibnamefont{Zenitani}},
  \bibinfo{journal}{Phys. Plasmas} \textbf{\bibinfo{volume}{14}},
  \bibinfo{pages}{112102} (\bibinfo{year}{2007}).

\bibitem[{\citenamefont{Zenitani}(2018)}]{zenitani18a}
\bibinfo{author}{\bibfnamefont{S.}~\bibnamefont{Zenitani}},
  \bibinfo{journal}{Plasma Phys. Control. Fusion}
  \textbf{\bibinfo{volume}{60}}, \bibinfo{pages}{014028}
  (\bibinfo{year}{2018}).

\bibitem[{\citenamefont{Zenitani and Hesse}(2008)}]{zenitani08a}
\bibinfo{author}{\bibfnamefont{S.}~\bibnamefont{Zenitani}} \bibnamefont{and}
  \bibinfo{author}{\bibfnamefont{M.}~\bibnamefont{Hesse}},
  \bibinfo{journal}{Phys. Plasmas} \textbf{\bibinfo{volume}{15}},
  \bibinfo{pages}{022101} (\bibinfo{year}{2008}).

\bibitem[{\citenamefont{Hesse et~al.}(2011)\citenamefont{Hesse, Neukirch,
  Schindler, Kuznetsova, and Zenitani}}]{hesse11a}
\bibinfo{author}{\bibfnamefont{M.}~\bibnamefont{Hesse}},
  \bibinfo{author}{\bibfnamefont{T.}~\bibnamefont{Neukirch}},
  \bibinfo{author}{\bibfnamefont{K.}~\bibnamefont{Schindler}},
  \bibinfo{author}{\bibfnamefont{M.}~\bibnamefont{Kuznetsova}},
  \bibnamefont{and} \bibinfo{author}{\bibfnamefont{S.}~\bibnamefont{Zenitani}},
  \bibinfo{journal}{Space Sci. Rev.} \textbf{\bibinfo{volume}{160}},
  \bibinfo{pages}{3} (\bibinfo{year}{2011}).

\bibitem[{\citenamefont{Petschek}(1964)}]{petschek64a}
\bibinfo{author}{\bibfnamefont{H.~E.} \bibnamefont{Petschek}}, in
  \emph{\bibinfo{booktitle}{Proc. AAS-NASA Symp. Phys. Solar Flares}}
  (\bibinfo{year}{1964}), vol.~\bibinfo{volume}{50} of
  \emph{\bibinfo{series}{NASA-SP}}, pp. \bibinfo{pages}{425--439}.

\bibitem[{\citenamefont{Lin and Lee}(1993)}]{lin93a}
\bibinfo{author}{\bibfnamefont{Y.}~\bibnamefont{Lin}} \bibnamefont{and}
  \bibinfo{author}{\bibfnamefont{L.~C.} \bibnamefont{Lee}},
  \bibinfo{journal}{Space Science Reviews} \textbf{\bibinfo{volume}{65}},
  \bibinfo{pages}{59} (\bibinfo{year}{1993}).

\bibitem[{\citenamefont{Levy et~al.}(1964)\citenamefont{Levy, Petschek, and
  Siscoe}}]{levy64a}
\bibinfo{author}{\bibfnamefont{R.~H.} \bibnamefont{Levy}},
  \bibinfo{author}{\bibfnamefont{H.~E.} \bibnamefont{Petschek}},
  \bibnamefont{and} \bibinfo{author}{\bibfnamefont{G.~L.}
  \bibnamefont{Siscoe}}, \bibinfo{journal}{AIAA J.}
  \textbf{\bibinfo{volume}{2}}, \bibinfo{pages}{2065} (\bibinfo{year}{1964}).

\bibitem[{\citenamefont{{Yi-Hsin Liu} et~al.}(2011)\citenamefont{{Yi-Hsin Liu},
  Drake, and Swisdak}}]{yhliu11b}
\bibinfo{author}{\bibnamefont{{Yi-Hsin Liu}}},
  \bibinfo{author}{\bibfnamefont{J.~F.} \bibnamefont{Drake}}, \bibnamefont{and}
  \bibinfo{author}{\bibfnamefont{M.}~\bibnamefont{Swisdak}},
  \bibinfo{journal}{Phys. Plasmas} \textbf{\bibinfo{volume}{18}},
  \bibinfo{eid}{092102} (\bibinfo{year}{2011}).

\bibitem[{\citenamefont{Mandt et~al.}(1994)\citenamefont{Mandt, Denton, and
  Drake}}]{mandt94a}
\bibinfo{author}{\bibfnamefont{M.~E.} \bibnamefont{Mandt}},
  \bibinfo{author}{\bibfnamefont{R.~E.} \bibnamefont{Denton}},
  \bibnamefont{and} \bibinfo{author}{\bibfnamefont{J.~F.} \bibnamefont{Drake}},
  \bibinfo{journal}{Geophys. Res. Lett.} \textbf{\bibinfo{volume}{21}},
  \bibinfo{pages}{73} (\bibinfo{year}{1994}).

\bibitem[{\citenamefont{Drake et~al.}(2008)\citenamefont{Drake, Shay, and
  Swisdak}}]{drake08a}
\bibinfo{author}{\bibfnamefont{J.~F.} \bibnamefont{Drake}},
  \bibinfo{author}{\bibfnamefont{M.~A.} \bibnamefont{Shay}}, \bibnamefont{and}
  \bibinfo{author}{\bibfnamefont{M.}~\bibnamefont{Swisdak}},
  \bibinfo{journal}{Phys. Plasmas} \textbf{\bibinfo{volume}{15}},
  \bibinfo{pages}{042306} (\bibinfo{year}{2008}).

\bibitem[{\citenamefont{Liu et~al.}(2018)\citenamefont{Liu, Hesse, Guo, Li, and
  Nakamura}}]{yhliu18c}
\bibinfo{author}{\bibfnamefont{Y.-H.} \bibnamefont{Liu}},
  \bibinfo{author}{\bibfnamefont{M.}~\bibnamefont{Hesse}},
  \bibinfo{author}{\bibfnamefont{F.}~\bibnamefont{Guo}},
  \bibinfo{author}{\bibfnamefont{H.}~\bibnamefont{Li}}, \bibnamefont{and}
  \bibinfo{author}{\bibfnamefont{T.~K.~M.} \bibnamefont{Nakamura}},
  \bibinfo{journal}{Phys. Plasmas} \textbf{\bibinfo{volume}{25}},
  \bibinfo{pages}{080701} (\bibinfo{year}{2018}).

\end{thebibliography}

\newpage

\end{document}